\documentclass[prl,twocolumn,superscriptaddress,nofootinbib]{revtex4}
\usepackage{graphicx}
\usepackage{dcolumn}
\usepackage{bm}
\usepackage{amsmath,amsthm,amssymb}

\begin{document}
%


\title{On asymptotic stability of the Skyrmion}

\author{Piotr Bizo\'n}
\affiliation{M. Smoluchowski Institute of Physics, Jagiellonian
University, Krak\'ow, Poland}
\author{Tadeusz Chmaj}
\affiliation{H. Niewodniczanski Institute of Nuclear
   Physics, Polish Academy of Sciences,  Krak\'ow, Poland}
   \affiliation{Cracow University of Technology, Krak\'ow,
    Poland}
\author{Andrzej Rostworowski}
\affiliation{M. Smoluchowski Institute of Physics, Jagiellonian
University, Krak\'ow, Poland}
\date{\today}
\begin{abstract}
We study the asymptotic behavior of spherically symmetric solutions
in the Skyrme model. We show that the
 relaxation to the degree-one soliton (called the Skyrmion) has a universal  form
 of a superposition of two effects:
  exponentially damped oscillations (the quasinormal ringing) and a
 power law decay (the tail). The quasinormal ringing, which dominates the dynamics for intermediate
 times,
  is a linear resonance effect. In contrast, the polynomial tail, which becomes uncovered at late times,
  is shown to
  be a \emph{nonlinear}
 phenomenon.

\end{abstract}

\maketitle

\emph{Introduction.} Stable stationary solutions are natural
candidates for the endstates of evolution of many physical systems.
The relaxation to these states is well understood for dissipative
systems, described by diffusion equations, however for conservative
Hamiltonian systems on unbounded domains the problem is much more
difficult because there is no local dissipation of energy and
convergence to equilibrium is due to dispersion, that is, radiation
of excess energy to infinity~\cite{w}. Understanding dissipation by
dispersion is important physically because the radiation emitted
during the approach to equilibrium  encodes information about the
attractor -- this kind of inverse problem has various applications,
for instance in identifying black holes via gravitation radiation
emitted during the last stages of gravitational collapse.

  In this paper we address the problem of relaxation to equilibrium in a
  very simple  setting of the
  spherically symmetric Skyrme model~\cite{ms}. This model, apart from its physical relevance in particle physics,
   is
  attractive theoretically
  because the equilibrium state is completely rigid: it has no moduli and
  no internal degrees of freedom, which makes the mathematical analysis feasible.
   Yet, despite the simplicity of the model, the relaxation process
  exhibits a surprising feature: after a
  transient oscillatory exponential decay (so called quasinormal ringing) which is a linear resonance effect,
  there proceeds a power law tail
  which has a nonlinear origin. Pointing out the failure
of the linear perturbation theory in capturing the asymptotic
dynamics is the main message of this paper.
  Below, after introducing the model, we first describe the quasinormal ringing using the
  linear perturbation theory,
  then we  present the numerical evidence for the asymptotic behavior of solutions,
  and finally we
  explain the tail using
   the nonlinear perturbation theory.
\vskip 0.2cm \emph{Background.} Let $M$ be a spacetime with a metric
$\eta_{\mu\nu}$ and $N$ be a complete Riemannian manifold with a
metric $g_{AB}$. Consider a map $U: M \rightarrow N$ and denote by
$S_{\mu\nu} = g_{AB}\,
\partial_{\mu} U^A \partial_{\nu} U^B$ the pulled back metric.
The (generalized) Skyrme model is defined by the lagrangian
\begin{equation}\label{L}
L = -\frac{1}{2} S_{\mu}^{\mu} + \frac{1}{4} \alpha^2 (S_{\mu\nu}
S^{\mu\nu} - S_{\mu}^{\mu} S_{\nu}^{\nu})\,,
\end{equation}
where $\alpha$ is the coupling constant having the dimension of
length. In this paper we consider the original Skyrme model~\cite{s}
where $M$ is the $3+1$ dimensional Minkowski spacetime with the
metric $\eta=-dt^2+dr^2+r^2 d\omega^2$ and $N$ is the 3-sphere with
the round metric $ds^2=dF^2+\sin^2{F} d\Omega^2$, where $d\omega^2$
and $d\Omega^2$ are the standard  metrics on the unit 2-sphere. We
restrict our attention to corotational maps for which  $F= F(t,r)$
and $\Omega=\omega$. For such maps the Euler-Lagrange equations
corresponding to (\ref{L})
 reduce to the single
nonlinear wave equation (using the abbreviation $w=r^2+2\alpha^2
\sin^2{F}$)
\begin{equation}\label{eq}
(w \dot F)^{\dot{} } - (w F')'
 + \sin(2F) + \alpha^2 \sin(2F) \left(\frac{\sin^2{F}}{r^2}\!+\! F'^2\!-\!\dot
 F^2\right)=0\,,
\end{equation}
where primes and dots denote derivatives with respect to $r$ and
$t$, respectively. We are interested in the long time behavior of
solutions of this equation for  smooth  finite energy initial data.
Regularity at the origin is ensured by the boundary condition
$F(t,0)=0$. The total conserved energy associated with solutions of
equation (\ref{eq}) can be written as the sum $E=E_{\sigma}+E_S$,
where
\begin{eqnarray}\label{energy}
E_{\sigma} &=&\frac{1}{2} \int_0^{\infty} \left[r^2({\dot F}^2
+{F'}^2) + 2\sin^2{F}\right]\: dr\,,\\ E_S&=&\frac{1}{2}\alpha^2
\int_0^{\infty} \left[\sin^2{F} ({\dot F}^2 +{F'}^2) +
\frac{\sin^2{F}}{r^2}\right] \: dr\,.
\end{eqnarray}
The quadratic part of the energy $E_{\sigma}$, corresponding to the
pure sigma model ($\alpha=0)$, has the supercritical scaling
$E_{\sigma}[F(x/\lambda)]=\lambda E_{\sigma}[F(x)]$, hence for
$\alpha=0$ it may be energetically favorable for solutions to shrink
and consequently singularities are expected to develop for some
initial data ~\cite{bct}. The quartic part of the energy $E_S$,
introduced by Skyrme, has the subcritical scaling
$E_S[F(x/\lambda)]=\lambda^{-1} E_S[F(x)]$, thus for nonzero
$\alpha$ shrinking of solutions to zero size is prevented by energy
conservation. Hereafter, we  assume that the coupling  constant
$\alpha$ is nonzero and use the unit of length such that $\alpha=1$.
 Note that for the total energy to be
  finite, solutions must satisfy the boundary condition
at spatial infinity $F(t,\infty)=m \pi$ ($m=0,1,\dots$), where an
integer $m$ is the topological degree of the map. Since the time
evolution is continuous (as long as no singularity forms), this
condition breaks the initial value problem into infinitely many
disjoint topological sectors labeled by the degree $m$. It is well
known that for each $m$ there is a unique regular static solution of
equation (\ref{eq})~\cite{lt}. Our numerical studies indicate that
these static solutions play the role of global attractors in the
evolution of regular corotational initial data of a given degree,
that is
 every solution starting
from smooth finite energy initial data of degree $m$ remains
globally regular for all times and asymptotically settles down to
the static solution  of degree $m$. Below, for concreteness, we
focus our attention on the degree-one sector $m=1$. In this case we
shall refer to the static solution as the Skyrmion and denote it by
$S(r)$. The Skyrmion is the most interesting corotational soliton
because, in contrast to solutions with $m>1$, it is stable with
respect to general (nonradial) perturbations. The profile function
$S(r)$ (see Fig.~1) is not known explicitly. The existence of the
Skyrmion was proved rigorously both by ODE techniques~\cite{lt} and
by variational methods~\cite{kl}.
\vskip 0.2cm \emph{Linear stability and quasinormal modes.}
In order to interpret the numerical results shown below we first
need to discuss the spectrum  the linear perturbations around the
Skyrmion.
 To this end we seek solutions in the form
$F(t,r)=S(r) + \delta F(t,r)$. Plugging this into  equation
(\ref{eq}), linearizing and using the auxiliary field $v(t,r)$
defined by
\begin{equation}
\delta F(t,r)=\frac{v(t,r)}{\sqrt{r^2+2 \sin^2{S}}}\,,
\end{equation}
we get the linear wave equation for the perturbation
\begin{equation}\label{leq}
\ddot v - v'' + \left(\frac{2}{r^2} + V \right) v =0\,,
\end{equation}
where the effective potential is
\begin{equation}\label{pot}
V=-4 a^2 \frac{1+3 a^2 + 3 a^4}{(1+2 a^2)^2}, \quad  \quad
a=\frac{\sin{S}}{r}\,.
\end{equation}
Near the origin $S(r)\sim b r$ (with $b\approx 2.0075$) and near
spatial infinity $S(r)\sim \pi -c/r^2$ (with $c\approx 2.1596$),
hence the potential $V(r)$ is finite at $r=0$ and falls off as
$r^{-6}$ for large $r$.
Substituting $v(t,r)=e^{-i k t} \psi(r)$ into equation (\ref{leq})
we obtain the $l=1$ radial Schr\"odinger equation
\begin{equation}\label{schr}
-\psi'' + \left(\frac{2}{r^2} + V(r)\right) \psi = k^2 \psi\,.
\end{equation}
It is known that this equation  has no bound states (and the
spectrum is purely continuous $k^2\geq 0$) which implies  that the
Skyrmion is linearly stable \cite{hds}.

Although linear stability is an important property of a soliton, it
provides little information about  the asymptotic behavior of
solutions near the soliton. The key concept in the studies of the
asymptotic stability of the soliton is the notion of the quasinormal
mode. The quasinormal mode (known also as the resonance) is a
regular solution of equation (\ref{schr}) which satisfies the
outgoing wave
 condition for $r \rightarrow \infty$
 \begin{equation}\label{out}
 \psi(r) \sim e^{i k r}, \quad k=\Omega - i \Gamma, \quad \Gamma>0.
 \end{equation}
The quasinormal mode with the least damping factor $\Gamma$ is
expected to dominate an intermediate stage of the relaxation to the
soliton. In order to find this mode we  use a shooting method which
goes as follows. First, we express $\psi$ in the amplitude-phase
form as $\psi=A \exp(i \phi)$ and rewrite equation (\ref{schr}) as
the following system
\begin{subequations}
\begin{equation}\label{eqA}
-A'' + A \phi'^2 +\left(\frac{2}{r^2} +V +\Gamma^2-\Omega^2\right)A
=0\,,
\end{equation}
\begin{equation}\label{eqphi}
 A \phi'' + 2 A' \phi' - 2 \Omega \Gamma A = 0\,.
\end{equation}
\end{subequations}
To ensure regularity at the center we require that
\begin{equation}\label{reg}
A(r) \sim r^2 \quad \mbox {and} \qquad \phi(r) \sim \frac{\Omega
\Gamma}{5} r^2 \quad \mbox {for} \quad r \rightarrow 0.
\end{equation}
We need to find $\Omega$ and $\Gamma$ such that the condition
(\ref{out}) is satisfied. A naive shooting method does not work
because an unwanted ingoing wave contamination of the condition
(\ref{out}) decreases exponentially with $r$ and cannot be tracked
numerically. To overcome this difficulty (which is intrinsic to the
problem and any numerical method has to cope with it)  we first
solve equations (\ref{eqA}) and (\ref{eqphi}) with the initial
condition (\ref{reg}) up to some relatively small intermediate
$r_0$.
Next, for $r>r_0$ we define the logarithmic derivative $
g=\psi'/\psi= A'/A + i \phi' $ and replace the Schr\"odinger
equation (\ref{schr}) by the Ricatti equation
\begin{equation}\label{ric}
g'+g^2-\frac{2}{r^2} - V + k^2=0\,.
\end{equation}
We solve this equation backwards in $r$ from some large $R$  to
$r_0$ starting with  the initial value
\begin{equation}
g(R) = \frac{k\, \hat h'_1(kR)}{\hat h_1(kR)},
\end{equation}
where the Ricatti-Hankel function $\hat h_1(kr)=(-i+\frac{1}{kr})
e^{i kr}$ is the exact outgoing wave solution of the free ($V=0$)
Ricatti equation (\ref{ric}). The value of $R$ should be chosen
sufficiently large so that the influence of the potential $V(R)$ is
negligible, however in practice $R$ should not be too large in order
to avoid numerical instabilities.  Matching the logarithmic
derivatives at the midpoint $r_0$ we found the quasinormal mode at
$k=0.61-0.26 i$.
\vskip 0.2cm \emph{Numerics.} We solved equation (\ref{eq})
numerically for different degree-one initial data and found that all
solutions remain globally regular and asymptotically settle down to
the Skyrmion (see Fig.~1).  Proving this asymptotic completeness
property is a challenging open problem which we do not pursue here
but take it as the starting point for further discussion.
\begin{figure}[h]
\centering
\includegraphics[width=0.49\textwidth]{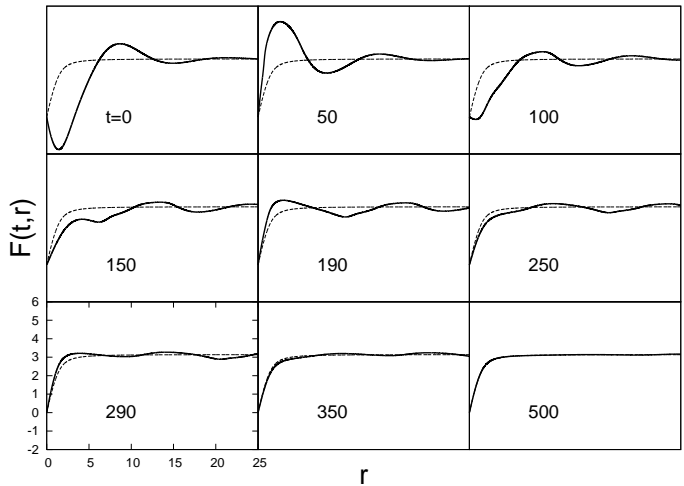}
\caption{\small{A  solution of degree one (solid line) is shown to
converge to the Skyrmion (dashed line).} }\label{fig1}
\end{figure}

Our aim is to understand the asymptotic dynamics of convergence to
the Skyrmion. Here we mean convergence in the pointwise sense, that
is we consider the behavior of a solution $F(t,r)$ for large $t$ at
a fixed distance $r$. The basic mechanism of decay is, of course,
dispersion - it is clearly seen in Fig.~1 how the excess energy is
being radiated away to infinity as the solution approaches the
Skyrmion. During the relaxation process one can distinguish two
universal stages of evolution: the oscillatory exponential decay
which we shall refer to as the quasinormal ringing and
 the polynomial decay which we shall refer to as the tail. The
 quasinormal ringing is a well-known  effect in the linear scattering
 theory~\cite{lp, ks}. As the name indicates, it is due to the presence of a quasinormal mode.
 The numerical confirmation of this fact is given in Fig.~2 where
 we show that for intermediate times the deviation of the solution from the Skyrmion
 is perfectly approximated by the fundamental quasinormal mode.
\begin{figure}[h]
\centering
\includegraphics[width=0.49\textwidth]{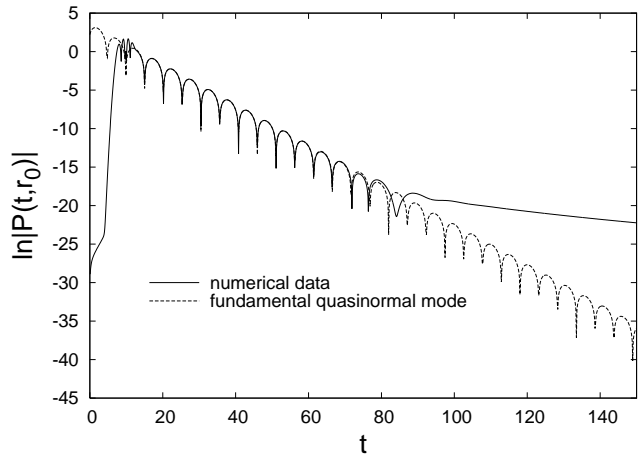}
\caption{\small{We plot $\ln|P(t,r_0=10)|$, where $P(t,r)=w \dot
F(t,r)$. Fitting the exponentially damped oscillation $P(t,r_0)=A
e^{-\Gamma t} \sin(\Omega t+\delta)$ to the numerical data on the
time interval $(20,60)$ we get $\Omega=0.610$ and $\Gamma=0.260$ in
perfect agreement with the perturbative calculation of the
fundamental quasinormal mode.} }\label{fig1}
\end{figure}

For later times the quasinormal mode becomes negligible and the
decay takes the form of a polynomial tail. The numerical computation
of the tail (see Fig.~3) gives  the power-law decay $F(t,r)-S(r)
\sim t^{-5}$.
\begin{figure}[h]
\centering
\includegraphics[width=0.49\textwidth]{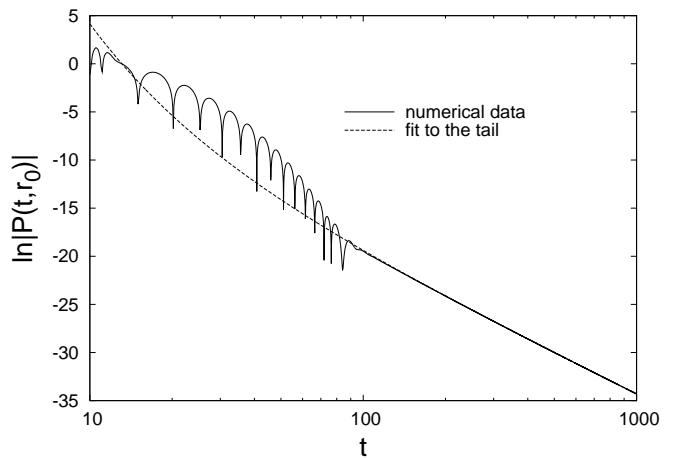}
\caption{\small{The same plot as in Fig.~2 but using the log-log
scale. Fitting the function $a-b\ln{t}+c/t$ (the power law decay
plus the first correction) to the numerical data on the interval
$(200,1000)$ we get $b=6.05$.}}\label{fig1}
\end{figure}
Surprisingly, the rate of decay of the tail is \emph{different} than
that predicted by the linear scattering theory. To see this, recall
that according to this theory \cite{clsy}, for compactly supported
initial data a solution of the linear radial wave equation with a
regular potential $V(r)$,
\begin{equation}\label{lin}
    \ddot v - v''+ \frac{l(l+1)}{r^2} v + V v =0\,,
\end{equation}
decays  at a fixed $r$ as $v(t,r) \sim t^{-\gamma}$, where
$\gamma=2l+\beta$ and $\beta>3$ is the rate of fall-off of the
potential at spatial infinity, i.e. $V(r)\sim r^{-\beta}$ for
$r\rightarrow \infty$ (for compactly supported or exponentially
localized potentials there are no  tails). In the case at hand
(equation (\ref{leq})) we have $l=1$ and $\beta=6$, hence according
to the linear theory the tail should have the power index $\gamma=8$
instead of the observed $\gamma=5$. Thus, the tail must be a
nonlinear effect and to understand it one needs to go beyond the
linear perturbation theory.  We found that the third order
perturbation theory provides a very good approximation of the tail.
Unfortunately, the perturbative expansion around the Skyrmion is
very messy and the technical details of the calculation might
obscure the key mechanism which is responsible for the nonlinear
tail. To avoid that, we will take advantage of the fact that the
same mechanism is operating in a simpler setting of the relaxation
to the vacuum in the topologically trivial sector $m=0$. In this
case there is no tail at all at the linear level.

  \vskip 0.2cm  \emph{Nonlinear
tail.} For topologically trivial initial data all solutions converge
asymptotically to the vacuum $F_0=0$. To determine the rate of
convergence, we substitute the
 expansion $F=\epsilon F_1+\epsilon^2 F_2 + \epsilon^3 F_3+
\mathcal{O}(\epsilon^4)$ into equation (\ref{eq}), where $\epsilon
F_1$ satisfies initial data while all $F_n$ with $n>1$ have zero
data. In the first order we get the free $l=1$ radial wave equation
\begin{equation}\label{f0}
\mathcal{L} F_1=0\,, \qquad \mathcal{L}:=\frac{\partial^2}{\partial
t^2}- \frac{1}{r^2}\frac{\partial}{\partial r}\left(r^2
\frac{\partial}{\partial r}\right)+ \frac{2}{r^2}\,,
\end{equation}
whose general regular solution  has the form
\begin{equation}\label{gf1}
    F_1(t,r)=\frac{a'(t-r)+a'(t+r)}{r}+\frac{a(t-r)-a(t+r)}{r^2}\,,
\end{equation}
where the function $a(r)$ is determined by initial data. We assume
that the initial data have compact support, hence $F_1$ has no tail
in agreement with  Huygens' principle.

In the second order $\mathcal{L} F_2=0$, hence $F_2$ vanishes, but
in the third order we get the inhomogeneous equation
\begin{equation}\label{f2}
\mathcal{L} F_3 = \frac{4}{3r^2} F_1^3+h\,,
\end{equation}
\begin{equation}\label{h}
h=\frac{2\alpha^2}{r^4}\left(F_1^3-2r F_1^2 F_1'+r^2 F_1
({F_1'}^2-{\dot F_1}^2)\right)\,.
\end{equation}
  We solve equation (\ref{f2}) using the retarded Green's function
  of the operator $\mathcal{L}$
\begin{equation}\label{green}
    G(t-t',r,r')=[|r-r'|\leq t-t'\leq r+r'] \, \frac{r^2+r'^2-(t-t')^2}{4
    r^2}\,.
\end{equation}
It follows from (\ref{gf1}) that for large $r$ the term $h$ is of
lower order in comparison with the first term on the right hand side
of equation (\ref{f2}), thus we can drop it without affecting the
leading order asymptotics of the tail. Then, using double null
coordinates $u=t'-r'$, $v=t'+r'$, we obtain
\begin{equation}\label{duh2}
    F_3(t,r)= \frac{2}{3 r^2} \int\limits_{|t-r|}^{t+r} dv
    \int\limits_{-v}^{t-r} \frac{(v-t)(t-u)+r^2}{(v-u)^2}
    F_1^3(u,v) du\,.
\end{equation}
We are interested in the asymptotic behavior of $F_3(t,r)$ for a
fixed $r$ and $t\rightarrow \infty$ (time-like infinity). Since the
initial data have compact support, in this limit we can change the
order of integration in (\ref{duh2}) and perform the integration
over $v$ explicitly. In the leading order we get
\begin{equation}\label{ntail}
    F_3(t,r) \sim c\, r\, t^{-5}\,,\qquad c=-\frac{64}{9} \int\limits_{-\infty}^{\infty}
    a'(u)^3
    du\,,
    \end{equation}
where the constant $c$ is the only trace of initial data. We have
verified numerically that this formula provides a very good
approximation of the tail for solutions having sufficiently small
initial data (see Fig.~4). \vskip 0.2cm \emph{Conclusions.} It
should be clear from the above discussion that the nonlinear tail is
not an exceptional feature of the Skyrme model but it is a general
phenomenon in scattering theory for nonlinear wave equations which
will be present whenever the backscattering due to an effective
potential around the attractor is weaker that the backscattering due
to a nonlinearity. This kind of phenomenon does not seem to have
been explored in the literature and we hope that our letter will
initiate investigations of an interplay between linear and nonlinear
effects in relaxation processes.
\begin{figure}[tbh]
\centering
\includegraphics[width=0.47\textwidth]{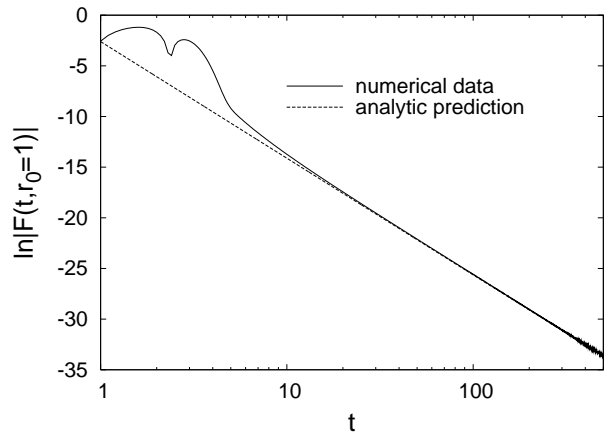}
\caption{\small{We plot the degree zero solution (solid line) for
initial data $F(0,r)=r^3 e^{-r^2}, \dot F(0,r)=0$ and superimpose
(dashed line) the analytic prediction for the tail (\ref{ntail})
with $c=35\sqrt{3\pi}/1458\approx 0.0737$ (note that there is no
adjustable parameter). The relative error (representing the
contribution from higher order iterations)  is $\sim 5\%$.
}}\label{fig1}
\end{figure}
\vskip 0.01cm \noindent {\emph{Acknowledgments:}} We thank Nikodem
Szpak for discussions. This research was supported in part by the
Polish Research Committee grant 1PO3B01229.

\end{document}